\documentclass[10pt,conference]{IEEEtran}
\IEEEoverridecommandlockouts

\usepackage{todonotes}
\usepackage{cite}
\usepackage{amsmath,amssymb,amsfonts}
\usepackage{algorithmic}
\usepackage{graphicx}
\usepackage{textcomp}
\usepackage{tabularx}
\usepackage{enumitem}
\usepackage{float}
\usepackage{multirow}
\usepackage{subcaption}
\usepackage{hyperref}
\usepackage{xcolor}
\usepackage{booktabs}
\usepackage[table]{xcolor}
\usepackage{tikz}
\usepackage{pgf-pie}
\usepackage{array}
\usepackage{pgfplots}

\pgfplotsset{compat=1.18}

\pgfplotsset{compat=1.18}

\definecolor{oiBlue}{HTML}{0072B2}
\definecolor{oiOrange}{HTML}{E69F00}
\definecolor{oiGreen}{HTML}{009E73}
\definecolor{oiYellow}{HTML}{F0E442}
\definecolor{oiSky}{HTML}{56B4E9}
\definecolor{oiVermillion}{HTML}{D55E00}
\definecolor{oiPurple}{HTML}{CC79A7}
\definecolor{oiGray}{HTML}{999999}
\definecolor{lightGray}{HTML}{D9D9D9}

\def\BibTeX{{\rm B\kern-.05em{\sc i\kern-.025em b}\kern-.08em
    T\kern-.1667em\lower.7ex\hbox{E}\kern-.125emX}}

\newcommand\changed[1]{\textcolor{black}{#1}}
\newcommand\changedd[1]{\textcolor{black}{#1}}

\begin{document}

\title{\changedd{Equal Access, Unequal Gains? GitHub Copilot and Productivity Among Developers with Disabilities}}

\newif\ifanonymous
\anonymousfalse 

\ifanonymous
\author{
\IEEEauthorblockN{Anonymous Authors}
\IEEEauthorblockA{}
}
\else
\author{
\IEEEauthorblockN{Emeralda Sesari}
\IEEEauthorblockA{
\textit{University of Groningen}\\
Netherlands\\
e.g.sesari@rug.nl}
\and
\IEEEauthorblockN{Daria Ana-Maria Mogoș}
\IEEEauthorblockA{
\textit{University of Groningen}\\
Netherlands\\
mogosdariaanamaria@gmail.com}
\and
\IEEEauthorblockN{Federica Sarro}
\IEEEauthorblockA{
\textit{University College London}\\
United Kingdom\\
f.sarro@ucl.ac.uk}
\and
\IEEEauthorblockN{Ayushi Rastogi}
\IEEEauthorblockA{
\textit{University of Groningen}\\
Netherlands\\
a.rastogi@rug.nl}
}
\fi

\maketitle

\begin{abstract}
Generative AI tools are increasingly used in software development and have been associated with benefits such as faster task completion and improved perceived productivity. However, whether these benefits extend across different developer populations remains unclear. This is especially relevant for developers with disabilities, whose work can already be impacted by accessibility barriers. Yet, evidence remains limited across a broader range of disability groups.
This study investigates whether the wider availability of GitHub Copilot is associated with changes in productivity among GitHub users who self-identify as having a disability. Using publicly available profile and repository data, we curated a dataset of 199 developers across multiple disability groups and compared their activity before and after GitHub Copilot became freely available. We found that most disability groups showed a descriptive pattern of fewer commits but more code modified. In several disability groups, the distribution of developers across commit activity levels shifted toward both lower and higher level, leaving fewer developers at moderate level. However, we found no evidence that changes in commits or net lines modified differed across disability groups. These findings suggest that the wider availability of GitHub Copilot cannot be associated with a broad increase or decrease in observable development activity, while patterns of change varied across developers and disability groups.
\end{abstract}


\section{Introduction}
Generative AI tools have become increasingly integrated into software development workflows. Tools such as GitHub Copilot can assist developers with tasks such as generating code, explaining code, writing documentation, and finding solutions to programming problems. Previous research suggests that AI-assisted development tools can influence developer productivity by accelerating programming tasks and improving developers' perceived productivity~\cite{peng2023the, barke2023grounded,10.1145/3520312.3534864}. \changed{However, existing studies on the productivity effects of AI have largely focused on the general developer population, and less is known about whether these benefits extend to other developer groups.}

Among these groups, developers with disabilities are particularly relevant when considering the effect of AI tools. Despite increasing attention towards inclusion and inclusive design in tech fields, accessibility efforts in software development have generally focused on end-user experience~\cite{10.1007/s10209-025-01210-w}. Accessibility initiatives often emphasize features such as screen reader compatibility, keyboard navigation, and accessible user interfaces to ensure products can be used by a wider range of individuals. Google's ``making accessible technology with and for people with disabilities" similarly focuses on integrating individuals with disabilities into the early testing cycles of product development with the purpose of improving the user experience~\cite{google_accessibility}. However, the needs and requirements of the developers themselves have received considerably less attention. 

According to the 2022 Stack Overflow Developer Survey~\cite{stackoverflow2022survey}, in which over 70,000 people participated, 1.7\% of respondents had a visual impairment, 0.85\% had a hearing impairment, and 0.44\% had a motor impairment. In addition to physical impairments, 10.57\% of the respondents identified they have a concentration and/or memory disorder, 10.31\% mentioned having an anxiety disorder, 9.71\% had a mood or emotional disorder, 4.27\% had autism, and 2.77\% had a learning difference (e.g., dyslexia)~\cite{stackoverflow2022survey}. 
Together, these figures show that developers with disabilities and neurodivergent conditions represent a non-negligible part of the developer population.

Developers with disabilities are particularly relevant when considering the potential benefits of Generative AI. 
For example, developers with motor impairments may benefit from reduced typing, while visually impaired developers may benefit from reduced navigation and easier access to information. 
Previous work has shown that neurodivergent developers can experience challenges related to task initiation, organization, focus, and context switching~\cite{gama2025sociotechnical}. Generative AI may provide new forms of support for neurodivergent users, for example by assisting with information access and emotional regulation~\cite{Glazko2025}. At the same time, these tools can introduce new accessibility challenges, and the benefits may differ depending on the developer and type of disability.

Among these tools, GitHub Copilot has received significant attention for its ability to generate code suggestions and provide conversational assistance~\cite{Solohubov2023AcceleratingSD}. \changed{GitHub has also positioned Copilot within its accessibility efforts as a tool that can help break down barriers to software development}~\cite{ziegler2022githubcopilotproductivity}, while previous studies have reported benefits related to task completion speed and perceived developer productivity~\cite{peng2023the,10.1145/3520312.3534864}. However, recent longitudinal evidence suggests that these perceived benefits do not necessarily translate into increased repository activity~\cite{stray2026developerproductivitygithubcopilot}.

Although previous research has started to examine Generative AI use among developers with disabilities, much of this work focuses on specific disability groups, particularly developers with visual impairments~\cite{cha2026gamechanger, Flores2025, chen2025screenreaderprogrammersvibe}. Less is known about whether the wider availability of AI-assisted development tools is associated with changes in the work of developers with disabilities across a broader range of disability groups. This is important because the benefits associated with AI tools may be particularly relevant to developers whose work is already affected by accessibility barriers.


In this paper, we investigate whether the wider availability of GitHub Copilot is associated with changes in the productivity of developers who publicly self-identify as having a disability on GitHub. Accordingly, we ask: \textbf{how has the observable development activity of self-identified disabled GitHub developers changed following the wider availability of GitHub Copilot?}.
From an initial pool of over 3,000 users, we curated usage data pertaining 199 developers across multiple disability groups based on publicly available profile information. Because actual adoption cannot be determined from repository data alone, we first examine explicit references to Generative AI tools in commit messages to understand the visibility of AI use within our sample. We then compare commits, net lines modified, and pull request acceptance rates before and after GitHub Copilot became freely available to all GitHub users. Through this analysis, we examine both overall changes in development activity and whether patterns of change differ across disability groups.
In the remaining of the paper we discuss background and related work (Section \ref{sec:related}), present the methodology we followed in our empirical study (Section \ref{sec:methodology}), report on the results (Section \ref{sec:Results}) and discuss their implications (Section \ref{sec:Discussion}) and future work (Section \ref{sec:Conclusions}).

\section{Background and Related Work} 
\label{sec:related}

\subsection{Generative AI and Developer Productivity}

Recent years have seen growing interest in understanding how AI-based code assistants may influence software development. Generative AI tools have attracted significant attention due to their widespread adoption and deep integration into modern development environments. These tools aim to help developers write code faster and with less effort, offering the opportunity to focus more on problem-solving and collaboration~\cite{githubcopilotdoc}. 

The 2026 study by Stray et al.~\cite{stray2026developerproductivitygithubcopilot} focused on measuring developer productivity and assessing the impact of Generative AI coding assistants in an organizational setting. The authors combined quantitative repository mining with self-reports and qualitative interviews within NAV IT, a Norwegian public-sector IT organization. 703 repositories were mined over a 2-year period, with six activity metrics used to measure productivity: number of commits per week, lines added per week, lines removed per week, net lines changed, commit frequency, and the change in weekly activity pre- and post-the wider availability of GenAI coding tools. 

They also examined perceived productivity through participants' subjective opinions, as well as qualitative metrics focused on reduced frustration, flow, and related measures ~\cite{stray2026developerproductivitygithubcopilot}. 
They found that tool users had committed more and changed more code than non-users; however, this pool was also more active than the other, even before the tool's availability. Tool adoption did not lead to a statistically significant increase in activity; however, participants reported feeling more productive, even though repository activity did not increase significantly. 


\subsection{Developers with Disabilities}
Research on developers with disabilities remains limited. Rocha et al.~\cite{10.1145/3674805.3690749} conducted a systematic literature review across multiple databases, focusing on papers published between 2013 and 2023. Only 39 papers explicitly addressed people with disabilities as members of software development teams, and only 11 discussed inclusion in depth. Research also appears to focus generally on a specific disability category: visual impairment, while neurodevelopmental, motor, and hearing impairments remain significantly less represented.

More recently, Gama et al.~\cite{gama2025sociotechnical} examined how Attention-deficit/Hyperactivity Disorder (ADHD) and Autism Spectrum Disorder (ASD) affect the work of software engineers. Their findings show that neurodivergent developers may experience challenges related to task initiation, organization and planning, focus, context switching, and emotional regulation, which can affect their software engineering performance. The study also highlights that accommodations and work practices can help regulate these challenges.

\subsection{Generative AI and Disability}
The focus on visual impairment is evident not only in broader research on developers with disabilities but also in disability-focused AI research.
Flores-Saviaga et al.~\cite{10.1145/3706598.3714008} conducted a study in 2025 that examined the impact of Generative AI coding assistants on developers who are visually impaired. The authors examined participants' usability, control, cognitive load, and workflow disruption and found that GitHub Copilot increased their sense of strategic control over the entire process, enabling them to focus on higher-level reasoning and design decisions. However, significant accessibility challenges were noted, including context switching. The participants also mentioned AI overload, when Copilot produced a larger number of suggestions in rapid succession, overwhelming users and increasing cognitive load. Despite these, participants had an overall optimistic outlook on the potential of AI coding assistants to improve accessibility, and the authors argued for activity-centered design principles specifically meant to account for non-visual workflows.

In 2025,  Microsoft~\cite{chen2025screenreaderprogrammersvibe} also conducted a study of programmers who use screen readers and AI code assistants, assessing how these tools empower software developers and the challenges they pose. This study involved analyzing recorded videos, transcripts, Copilot chat logs, and participant-shared records in a "diary" format. Throughout the study, respondents consistently perceived value in using GitHub Copilot, as all reported that their programming efficiency was enhanced by it.

Cha et al.~\cite{Cha} conducted a study in 2025, focusing on the use of Generative AI by software professionals with visual impairments in the workplace. The study was qualitative, using interviews to assess how participants felt about Generative AI tools such as GitHub Copilot, Claude, Gemini and DeepSeek, in a task-dependent manner. They found that Generative AI reduced access labor, as there appeared to be less reliance on sighted colleagues, and reduced the need to build do-it-yourself (DIY) accessibility tools accessibility tools for participants' specific needs. However, the possibility of hallucinations remained as a risk, which for users with vision difficulties might be harder to detect, increased vulnerability due to errors going unnoticed, and, most common of all studies presented so far, the incompatibility with screen readers. 

While visual impairment seemed to be a primary focus of disability-centered AI research, substantially less work has examined other disability categories. Mapping this research focus is important for assessing whether the benefits of AI-driven accessibility tools are distributed evenly across disability groups or whether certain groups are underrepresented in design and evaluation efforts.

Although Huffman et al.~\cite{huffman2025weuseithearing} did not focus on developer productivity, they examined how individuals in the Deaf and Hard of Hearing (DHH) community use large language models. A key finding was that participants often preferred AI over human tutors due to increased privacy and reduced embarrassment during learning. However, experiences varied by language preference: American Sign Language (ASL)-primary users reported increased comfort, while some English-primary users found AI responses unnatural and less comfortable. A major challenge for participants was difficulty writing effective prompts, particularly for ASL-primary users, leading to suggestions for ASL-aware support, such as training LLMs on ASL gloss. Participants also highlighted a lack of DHH-specific AI education and training resources, as most existing guidance assumes hearing users.

Pivoting to non-physical disabilities, Glazko et al. ~\cite{Glazko2025} conducted a study on how neurodivergent users use Generative AI. Their results highlighted that Generative AI was used to reduce executive dysfunction load, manage fatigue, and handle brain fog. Generative AI was used to start tasks, organize thoughts, and manage ``task paralysis", but one of the primary uses was to mask neurodivergent traits and meet normative communication expectations. 

Similarly, Saraswat et al.~\cite{article} conducted a study in 2024 on how AI can be used to assist people with neurodiversity. The paper argues that AI can remove repetitive, low-engagement work, allow neurodivergent individuals to focus on their strengths, and improve job matching. They highlighted that AI systems are trained on neurotypical data, which may distort decision-making, exclude neurodivergent cognitive styles, and may reinforce stereotypes. They also pointed out that GAI interfaces may be sensory-overstimulating for neurodivergent people, with rapid chat outputs, ambiguous instructions, and cognitive overload from excessive options. 

These studies show that disability is not a monolithic category, but rather encompasses multiple dimensions and types. 
Even within a single category, such as neurodivergence, diversity exists. Treating neurodivergent developers as a homogeneous group may obscure important differences in cognitive profiles and interaction patterns. 
Nashiha et al~\cite{Inklusivitas} classify people with disabilities into four categories: \emph{Physical} disabilities, \emph{Sensory} disabilities, \emph{Intellectual} disabilities and \emph{Mental} disabilities.

In a nutshell, literature suggests that Generative AI tools may influence how people work, affecting both measurable activity metrics and users' perceptions of their productivity. However, little is known about how AI-assisted development tools impact developer productivity across a broader range of disability categories.

\section{Methodology}
\label{sec:methodology}

To investigate whether the wider availability of Generative AI tools was associated with changes in observable productivity-related activity among developers with disabilities, we used publicly available GitHub data. Since our goal was to examine productivity changes among a specific population, that of developers who self-identify as having disabilities, we opted for a longitudinal approach that compares activity before and after a specific date. This date was selected based on the moment when GitHub Copilot became available to all GitHub users, regardless of their tiers/subscription plans. This methodology allowed us to identify relevant developers, define productivity-related measures, and examine whether the activity changed after the wider availability of GenAI tools.

\subsection{Selection of GitHub Copilot}
As stated in the previous section, GitHub Copilot has shown a positive impact on productivity, in terms of both developers' happiness and speed~\cite {githubcopilotprod}, while users had access only to the preview version of the tool. GitHub's popularity also motivates our choice for GitHub Copilot, as a large percentage of software developers use this platform~\cite{articleGit}. However, now that Copilot has been available for free to all GitHub tiers for over a year (18 December 2024)~\cite{githubcopilotfree}, we believe it may be safe to assume that the majority of engineers working in software development have at least heard of this tool (or similar ones), if they have not used it yet. We therefore use the date on which Copilot became freely available as the dividing point between the pre-availability and post-availability periods. 

Building on related work in Section~\ref{sec:related}, we treat productivity as a multidimensional concept that cannot be fully captured by repository activity alone. In this study, we therefore use commits, net lines modified, and pull request acceptance rate as observable indicators of development activity rather than as complete measures of productivity. GitHub Copilot was selected as the focal GenAI tool because it has been widely studied in software development and has been examined using both repository-based and self-reported measures of developer productivity. This allows our study to build on existing work while extending the analysis to developers with disabilities, who remain underrepresented in prior research.

\subsection{Defining Productivity}

Productivity emerges from the interaction of multiple aspects, which include satisfaction and well-being, performance outcomes, developer activity, collaboration, and efficiency (SPACE)~\cite{Forsgren2021}. Complementing this perspective, frameworks such as DORA focus on measurable delivery performance and system reliability, including deployment frequency, change lead times, change failure rates, and time to restore service. Additionally, the HEART framework measures happiness, engagement, adoption, retention, and task success, viewing productivity from a user-centered perspective~\cite{unknown}. 

There appears to be consensus in the literature that productivity cannot be accurately measured, as many factors influence how productive an individual is, depending on the criteria used. Therefore, for the purposes of this study, we first need to define appropriate metrics for measuring productivity.
\begin{itemize}
    \item \textit{Number of Commits Before and After~\cite{stray2026developerproductivitygithubcopilot}}. We used this metric to measure the frequency of contributions. Because GitHub activity is highly skewed, we used the median rather than the mean. This was the case for all metrics used in our analysis.
    \item \textit{Net Lines Modified Before and After(Lines Added + Removed)~\cite{stray2026developerproductivitygithubcopilot}}. Net lines modified is defined as the sum of lines added and lines removed, rather than the difference, because lines removed still signify activity. This measured how much code work was done. We choose this as a complementary metric because previous studies have found that different demographics exhibit distinct activity patterns~\cite{terrell}.
    \item \textit{Acceptance Rate (number of PRs Merged/ number of PRs created) Before and After}. Activity alone does not always indicate a change in productivity if the quality of the work has decreased while the number of commits has increased. This is why we also chose to examine the acceptance rate of PRs merged during the before-and-after periods, as merged PRs indicate that the work is of acceptable quality, especially in collaborative projects~\cite{LENARDUZZI2021110806}.
\end{itemize}


\subsection{Participant Selection}
We aimed to conduct a more general study than previous studies. For example, hearing impairment is not binary; it comes in degrees. These range as: slight loss/minimal loss (hearing limited to 16-25 dB), mild loss (hearing limited to 26-40 dB), moderate loss (hearing limited to 41-55 dB), moderate/severe loss (hearing limited to 56-70 dB), severe loss (hearing limited to 71-90 dB) and profound loss (hearing limited to 91 dB or more)~\cite{deafandblindoutreach2026}. However, it might not be likely that GitHub users will use this level of detail in their profile descriptions. Therefore, we have curated the keyword list for this disability in an attempt to be comprehensive, based on both official definitions~\cite{deafandblindoutreach2026} and Reddit posts~\cite{reddit_deaf_terminology_2023} related to hearing impairments as shown in Table~\ref {tab:keywords}

For the neurodivergent category, this is conceptually defined in research as a range of neurodevelopmental differences that are recognized both clinically and academically. Accordingly, we curated the keyword list in Table \ref{tab:keywords}~\cite{Jepson2024}.
We also considered a more general umbrella term, "disabled," to encompass a broader user base, as many users do not prefer to disclose their specific disabilities in their GitHub profile descriptions.

We also included developers with visual impairments to broaden the range of disabilities represented in our sample, particularly because prior work and accessibility efforts have frequently focused on visual impairment~\cite{github_copilot_screen_reader_guide}.  The keywords used to identify these developers are also shown in Table~\ref{tab:keywords}.


\subsection{Data Collection}
We started by identifying GitHub users whose profile descriptions included any of the keywords listed in Table~\ref{tab:keywords}.
\begin{table}[t]
\caption{Disability Keywords}

\label{tab:keywords}
\centering
\footnotesize

\begin{tabularx}{\columnwidth}{p{0.23\columnwidth}|X}
\hline
\textbf{Category} & \textbf{Keywords} \\
\hline

\textbf{Physical} &
deaf, hard of hearing, slight hearing loss, minimal hearing loss, mild hearing loss, moderate hearing loss, severe hearing loss, profound hearing loss \\
\hline

\textbf{Non-Physical} &
dyscalculia, dyslexia, attention deficit disorder, ADHD, ADD, Tourette's syndrome, dyspraxia, developmental coordination disorder, autism, ASD, AuDHD \\
\hline

\textbf{General} &
disabled \\
\hline

\textbf{Visual} &
visually impaired, blind, low vision, partially blind, partially sighted,
legally blind, sight loss, BLV, BVI \\
\hline
\end{tabularx}
\end{table}
To account for GitHub's search result limits, we stratified our searches by follower count, following prior work that suggests an association between follower count and developer activity~\cite{Liao}. From this mining script, we gathered a total of 9753 users. To ensure the keywords mined from these profiles are standalone words rather than part of other words, we added an additional round of checks. This left us with a total of 3163 users.

Next, we wanted to select users with representative activity levels across the two time frames. Many users had very low or zero activity, so to properly remove them, we computed a lower activity threshold for the number of commits using the 25th percentile. From this, our lower bound for the before period was at least 7 commits and for the after period, 5 commits. This yielded a dataset of 543 users with sufficient activity. 

\begin{table}[t]
\caption{File Extensions}
\label{tab:file_extensions}
\centering

\begin{tabularx}{\columnwidth}{p{0.25\columnwidth}|X}
\hline
\textbf{Category} & \textbf{Extensions} \\
\hline

Programming Languages &
Python, JavaScript, TypeScript, Java, C, C++, C\#, Go, Rust, Ruby, PHP, Kotlin, Swift, Scala, Dart \\
\hline

\end{tabularx}
\end{table}

We then validated whether these users were software developers. The purpose of our study was to assess the productivity of software developers with disabilities, with varying levels of activity. We did not want to assess only professionals, but also individuals who are software developers even as a hobby. However, we were not interested in users who use GitHub for other purposes, such as writers, designers, or people hosting notes. Therefore, we checked users' repositories for the file extensions listed in Table~\ref{tab:file_extensions}. This removed 20 users, leaving 523 users.

After several rounds of review, we found that many of our keywords were overly specific and \changed{matched only one developer}. However, words such as ``ADD" could pose a threat, as even when they were standalone words, they might refer to another meaning instead of ``attention deficit disorder". To ensure the validity of our users, we conducted a manual check. ADD represented 56\% of our population, meaning 297 users. Checking all of these users manually would be very time-consuming and inefficient. Therefore, we chose to use sampling. 

Initially, we examined the first 31 users (more than 10\% of the population) in our dataset who had the keyword ``add" in their profile descriptions. It turned out 100\% of these users used the word ``add" with another meaning, most of which had some sort of ``Add a bio" left in their profile descriptions or the only text present in their profile descriptions. To add more rigor, we analyzed 20 additional users for stronger validation. Of the 51 users we looked at in this sample, 100\% did not mean they had ADD. Due to this 100\% false-positive rate, the keyword was removed from the analysis. This left us with a total of 226 users.

Using the same profile-based search and validation procedure, we also identified developers using the visual-impairment keywords listed in Table~\ref{tab:keywords}. After removing low-activity users, using the 25th percentile again, and ensuring we considered only software developers based on their file extensions, we had 43 sufficiently active visually impaired users. For the pre-availability period, the minimum activity requirement was 7 commits; for the post-availability period, it was 8 commits. This did not affect the threshold for the other categories. Combining these users with the previous sample resulted in 269 unique and sufficiently active developers.

\subsubsection{Manually Validating Disabilities}
\changed{After removing the ``ADD" keyword, we separately checked all 269 remaining profiles to confirm that the disability terms were used to describe the developers themselves.} Words such as   ``blind" and ``deaf" are often used with different meanings, depending on the context. As such, users who, for example, may write in their profile descriptions   ``Companies are deaf and blind" do not self-identify as disabled. To ensure our analysis is performed solely on users with disabilities, we had to manually check each user's bio. The removal of a user was determined through natural language processing. If a bio contained only the term   ``disability" itself, the user was categorized as having that disability. If a bio used any disability to refer to other entities, such as   ``companies are blind[...]", or in a speculative manner:   ``potentially ADHD", or mentioned working with people with disabilities, the respective users were removed from our dataset.

After manually checking all 269 developers, we were left with 199 disabled developers with sufficient activity for our analysis. Developers whose profiles did not show a disability disclosure
were kept as a comparison sample (i.e., comparison group). These developers came from
the same data collection process, but they were not selected to
match the disability sample.

\subsubsection{Identifying Explicit AI Mentions}
Identifying actual GenAI use from repository data is difficult because developers do not necessarily document whether AI contributed to their code. We therefore examined commit messages for explicit references to GenAI tools or AI-assisted code generation. As such, we have mined commit messages and looked at those that mention AI use. Initially, we only looked at commit messages that mentioned Copilot or GitHub Copilot. \changed{As relatively few commit messages explicitly mentioned Copilot, we broadened the search by looking} at other Gen AI tools. We have compiled a list of keywords and formed larger groups of tool categories, as shown in Table \ref{tab:ai_keywords}.
\changed{We then summarized these mentions, in Section \ref{visibility}, to describe how visible Generative AI use was in the repository activity of our sample.}

\begin{table}[t]
\caption{AI Categories in Commit Messages}
\label{tab:ai_keywords}
\centering

\begin{tabularx}{\columnwidth}{lX}
\hline
\textbf{Category} & \textbf{Keywords} \\
\hline
Copilot & copilot, GitHub Copilot \\
ChatGPT & chatgpt, openai \\
Claude & claude, anthropic \\
Gemini & gemini, bard \\
Bing / Microsoft AI & bing, microsoft ai \\
Grok / xAI & grok, xai \\
Other AI Tools & codewhisperer, tabnine, ghostwriter, replit ai \\
General AI Mentions & ai-generated, generated by ai, ai-assisted, llm, large language model, using ai, written by ai \\
\hline
\end{tabularx}
\end{table}

\subsection{Data Analysis}

The scope of our study was to examine whether the activity levels of developers with disabilities whose data is publicly available on GitHub have changed, with the incipient moment defined as the date GitHub Copilot became available for free to all GitHub users on 18 December 2024. For that, we performed our analysis on two levels.

\subsubsection{First Level - Category-Based Analysis}
We first grouped developers into four mutually exclusive disability categories:
\begin{enumerate}
    \item \textbf{Physical Disabilities}: ``deaf";
    \item \textbf{Non-Physical}: ``dyscalculia", ``dyslexia", ``adhd", ``autism", ``asd", ``audhd";
    \item \textbf{Visual}: ``visually impaired", ``blind", ``legally blind";
    \item \textbf{Disabled}: ``disabled".
\end{enumerate}

We kept developers who used the general term "disabled" as a separate category because this label alone does not provide sufficient information to classify a developer as having a physical or non-physical disability. This avoided including the same developers in multiple broader disability categories. The resulting categories contained 13 Physical, 121 Non-Physical, 20 Visual, and 45 Disabled developers. We also included the comparison sample in the between-group analysis.

For each developer, we calculated a change score for each activity metric by subtracting the pre-availability value from the post-availability value. Here, we ask if the broader groups differ in how much their activity changed from before to after the availability of Copilot. We used the \textsc{Kruskal-Wallis H}~\cite{kruskal1952} test to examine whether these changes differed across the four disability categories and the comparison group (i.e., developers with no identified disability disclosure). We additionally conducted exploratory pairwise comparisons using the \textsc{Mann-Whitney U} test~\cite{mann1947}, with \textsc{Cliff's delta}~\cite{cliff1993} reported as an effect size. Since multiple pairwise comparisons were conducted, their p-values were adjusted using the Benjamini-Hochberg procedure~\cite{benjamini1995} to control the false discovery rate.

\subsubsection{Within-Group and Within-Category Analysis}
To evaluate whether activity changed significantly within each disability group, we applied the \textsc{Wilcoxon signed-rank} test~\cite{wilcoxon1945}. This test compares paired observations for the same developers before and after the wider availability of GitHub Copilot. For example, if ADHD developers' commit changed after wider Copilot availability. This test was selected because of the paired nature and non-normal distribution of the activity measures. Effect size was reported using the standardized \textsc{Wilcoxon r statistic}\changed{~\cite{hollander2013nonparametric,bauer1972constructing}}.

We also examined whether the magnitude of change differed between disability subtypes belonging to the same broader category. For example, if ADHD, Autism, ASD, and AuDHD differ in how much their commits changed from before to after availability. In this analysis, only subtypes containing at least 10 developers were included. After defining the broader categories as mutually exclusive, only the Non-Physical category contained multiple eligible subtypes: ADHD, Autism, ASD, and AuDHD. We therefore used the \textsc{Kruskal-Wallis H} test to compare change scores across these subtypes.

\subsubsection{Minimal group size}\label{sec:min_group_size}
We only performed statistical tests for disability subgroups with at least \textit{10 users}. We used this threshold to avoid drawing conclusions from very small groups, while still keeping as many disability groups as possible in the analysis.
For pull request acceptance rate, we additionally required developers to have a non-zero acceptance rate in both periods. After applying this requirement, no broader group had at least 10 eligible developers. We therefore report only the descriptive of acceptance-rate changes.

\section{Results}
\label{sec:Results}
\subsection{Data Distribution} 
\label{data_distribution}

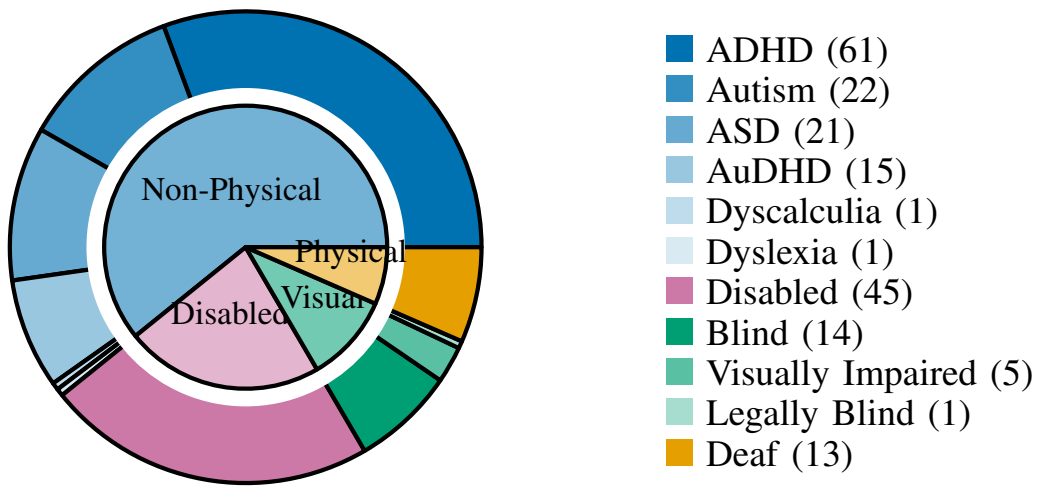
\begin{figure}[t]
\centering

\makebox[\columnwidth][c]{%
\begin{minipage}[c]{0.36\columnwidth}
\centering
\begin{tikzpicture}[scale=0.78, every node/.style={transform shape}]

\pie[
    sum=199,
    radius=2,
    hide number,
    color={
        oiBlue!100, oiBlue!80, oiBlue!60, oiBlue!40, oiBlue!25, oiBlue!15,
        oiPurple!100,
        oiGreen!100, oiGreen!65, oiGreen!35,
        oiOrange!100
    }
]{
    61/,
    22/,
    21/,
    15/,
    1/,
    1/,
    45/,
    14/,
    5/,
    1/,
    13/
}

\fill[white] (0,0) circle (1.35);

\pie[
    sum=199,
    radius=1.2,
    text=inside,
    hide number,
    font=\footnotesize,
    color={
        oiBlue!55,
        oiPurple!55,
        oiGreen!55,
        oiOrange!55
    }
]{
    121/Non-Physical,
    45/Disabled,
    20/Visual,
    13/Physical
}

\end{tikzpicture}
\end{minipage}
\hspace{0.03\columnwidth}
\begin{minipage}[c]{0.46\columnwidth}
\centering
\scriptsize
\renewcommand{\arraystretch}{0.95}
\begin{tabular}{@{}l@{}}
\textcolor{oiBlue!100}{\rule{5pt}{5pt}} ADHD (61) \\
\textcolor{oiBlue!80}{\rule{5pt}{5pt}} Autism (22) \\
\textcolor{oiBlue!60}{\rule{5pt}{5pt}} ASD (21) \\
\textcolor{oiBlue!40}{\rule{5pt}{5pt}} AuDHD (15) \\
\textcolor{oiBlue!25}{\rule{5pt}{5pt}} Dyscalculia (1) \\
\textcolor{oiBlue!15}{\rule{5pt}{5pt}} Dyslexia (1) \\
\textcolor{oiPurple!100}{\rule{5pt}{5pt}} Disabled (45) \\
\textcolor{oiGreen!100}{\rule{5pt}{5pt}} Blind (14) \\
\textcolor{oiGreen!65}{\rule{5pt}{5pt}} Visually Impaired (5) \\
\textcolor{oiGreen!35}{\rule{5pt}{5pt}} Legally Blind (1) \\
\textcolor{oiOrange!100}{\rule{5pt}{5pt}} Deaf (13) \\
\end{tabular}
\end{minipage}%
}
\caption{Nested distribution of disability groups. The inner ring
shows broader categories and the outer ring shows individual groups.}
\label{fig:disability_distribution}
\end{figure}

As illustrated in Figure~\ref{fig:disability_distribution}, the largest disability category in our study is the Non-Physical category. As explained in Section~III-E, the four disability categories were defined as mutually exclusive. The disability sample consisted of 121 Non-Physical, 45 Disabled, 20 Visual, and 13 Physical developers. In addition, the comparison group contained 70 unique developers for whom no disability disclosure was identified.


As shown in Figure ~\ref{fig:disability_distribution}, ADHD is the largest group, representing 30.7\% (61 users), followed by disabled (22.6\%, 45 users), Autism (11.1\%, 22 users), and ASD (10.6\%, 21 users). AuDHD (7.5\%, 15 users), blind (7.0
deaf (6.5\%, 13 users) are also somewhat well represented.
But the remaining groups of visually impaired (2.5\%, 5 users),
dyscalculia (0.5\%, 1 user), dyslexia (0.5\%, 1 user), and legally
blind (0.5\%, 1 user) have small representation in the dataset.


\begin{figure}[t]
\centering



\begin{subfigure}[t]{0.48\columnwidth}
\centering

\begin{tikzpicture}
\pie[
    sum=58,
    radius=1.35,
    text=pin,
    hide number,
    font=\footnotesize,
    color={
        oiBlue,
        oiOrange,
        oiVermillion,
        oiPurple,
        oiGreen,
        oiSky,
        oiYellow,
        oiGray}
]{
    16/16,
    16/16,
    8/8,
    8/8,
    7/7,
    2/2,
    1/1
}

\fill[white] (0,0) circle (0.52);

\node[align=center] at (0,0)
    {\footnotesize\textbf{58}\\[-1pt]\scriptsize mentions};

\end{tikzpicture}

\caption{Disability disclosures}
\label{fig:tool_distribution_disabled}
\end{subfigure}
\hfill
\begin{subfigure}[t]{0.48\columnwidth}
\centering

\begin{tikzpicture}
\pie[
    sum=25,
    radius=1.35,
    text=pin,
    hide number,
    font=\footnotesize,
    color={
        oiBlue,        
        oiOrange,      
        oiVermillion,  
        oiPurple,      
        oiGreen,       
        oiSky,         
        oiYellow,      
        oiGray}        
]{
    6/6,  
    4/4,  
    3/3,  
    5/5,  
    5/5,  
    1/1,  
    1/1   
}

\fill[white] (0,0) circle (0.52);

\node[align=center] at (0,0)
    {\footnotesize\textbf{25}\\[-1pt]\scriptsize mentions};

\end{tikzpicture}
\caption{Comparison sample}
\label{fig:tool_distribution_comparison_sample}
\end{subfigure}

\vspace{3pt}


\footnotesize
\renewcommand{\arraystretch}{1.1}
\setlength{\tabcolsep}{3pt}

\begin{tabular}{@{}llll@{}}

\textcolor{oiBlue}{\rule{6pt}{6pt}} ChatGPT
&
\textcolor{oiOrange}{\rule{6pt}{6pt}} Claude
&
\textcolor{oiVermillion}{\rule{6pt}{6pt}} Gemini
&
\textcolor{oiPurple}{\rule{6pt}{6pt}} General AI
\\

\textcolor{oiGreen}{\rule{6pt}{6pt}} Copilot
&
\textcolor{oiSky}{\rule{6pt}{6pt}} Bing
&
\textcolor{oiYellow}{\rule{6pt}{6pt}} Grok
&
\textcolor{oiGray}{\rule{6pt}{6pt}} Other tools

\end{tabular}

\caption{Distribution of AI tool mentions in commit messages for developers
with and without identified disability disclosures.}
\label{fig:tool_distribution_comparison}

\end{figure}
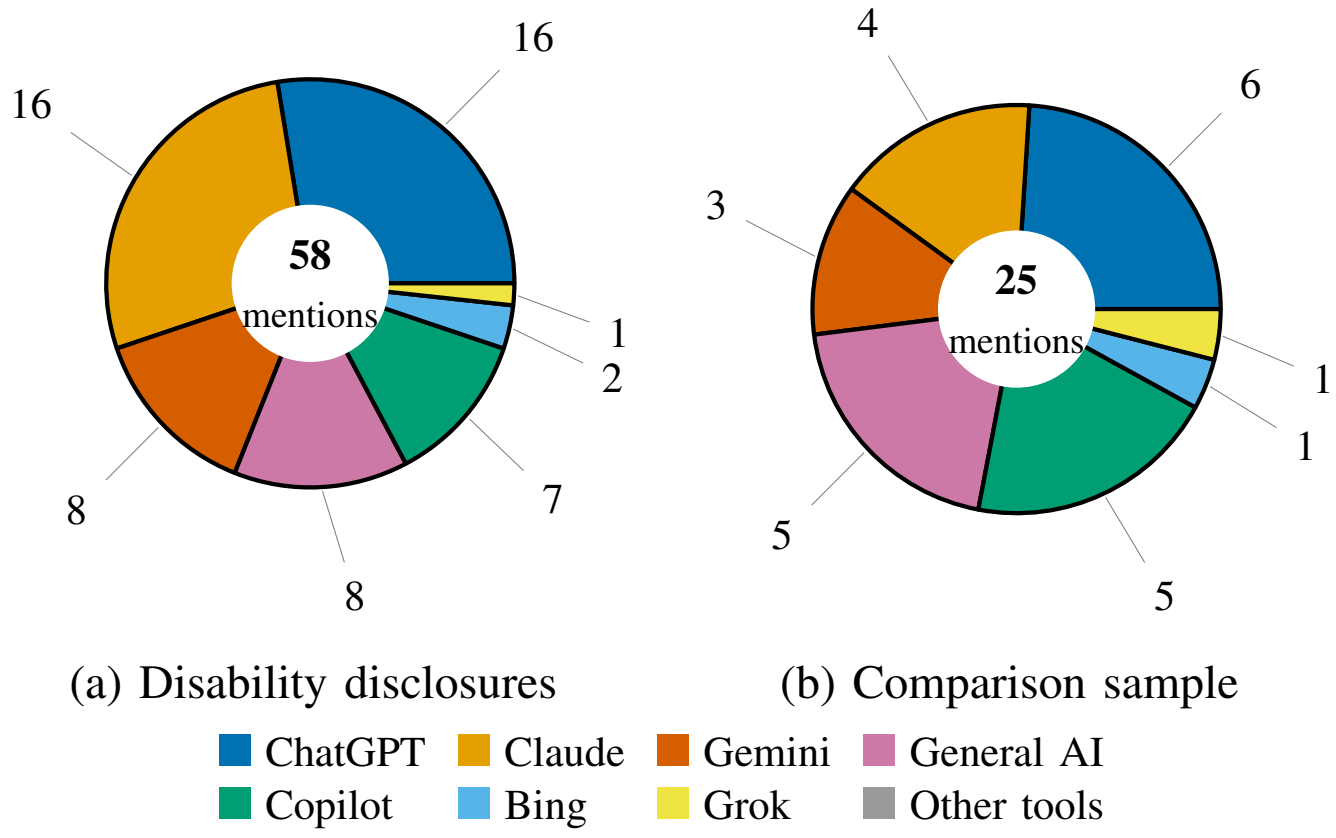

\subsection{Visibility of Generative AI in Developer Activity}\label{visibility}
Before examining changes in productivity, we first looked at whether Generative AI use was visible in the commit activity of developers in our sample.
Figure~\ref{fig:tool_distribution_comparison} shows the distribution of AI tools explicitly mentioned in commit messages by developers with and without identified disability disclosures. Among developers with disabilities, ChatGPT and Claude were the most frequently mentioned tools, each accounting for 27.6\% of the identified mentions. Among developers in the comparison sample, ChatGPT was the most frequently mentioned tool (24\%), followed by general AI mentions and Copilot, each accounting for 20\% of the identified mentions.

We then examined how many developers had at least one explicit reference to a Generative AI tool in their commit messages. Overall, 17\% of developers with disability disclosures had at least one explicit AI-related mention, compared with 20\% of developers in the comparison sample.
These percentages represent only AI use that was explicitly documented in commit messages and should not be interpreted as adoption rates. Developers are not generally required to disclose the use of Generative AI tools when committing code, and therefore actual use may be higher than what is visible through these mentions. \changed{In this analysis, we found similar proportions of explicit AI-related mentions across the two samples, which suggest that Generative AI use was visible at broadly \changedd{similar levels as the developers without disability} in their repository activity.}

Having found explicit Generative AI mention in the repository activity of some developers in our sample, we next examine whether developers' productivity changed following the wider availability of GitHub Copilot.

\subsection{Changes in Productivity Metrics}



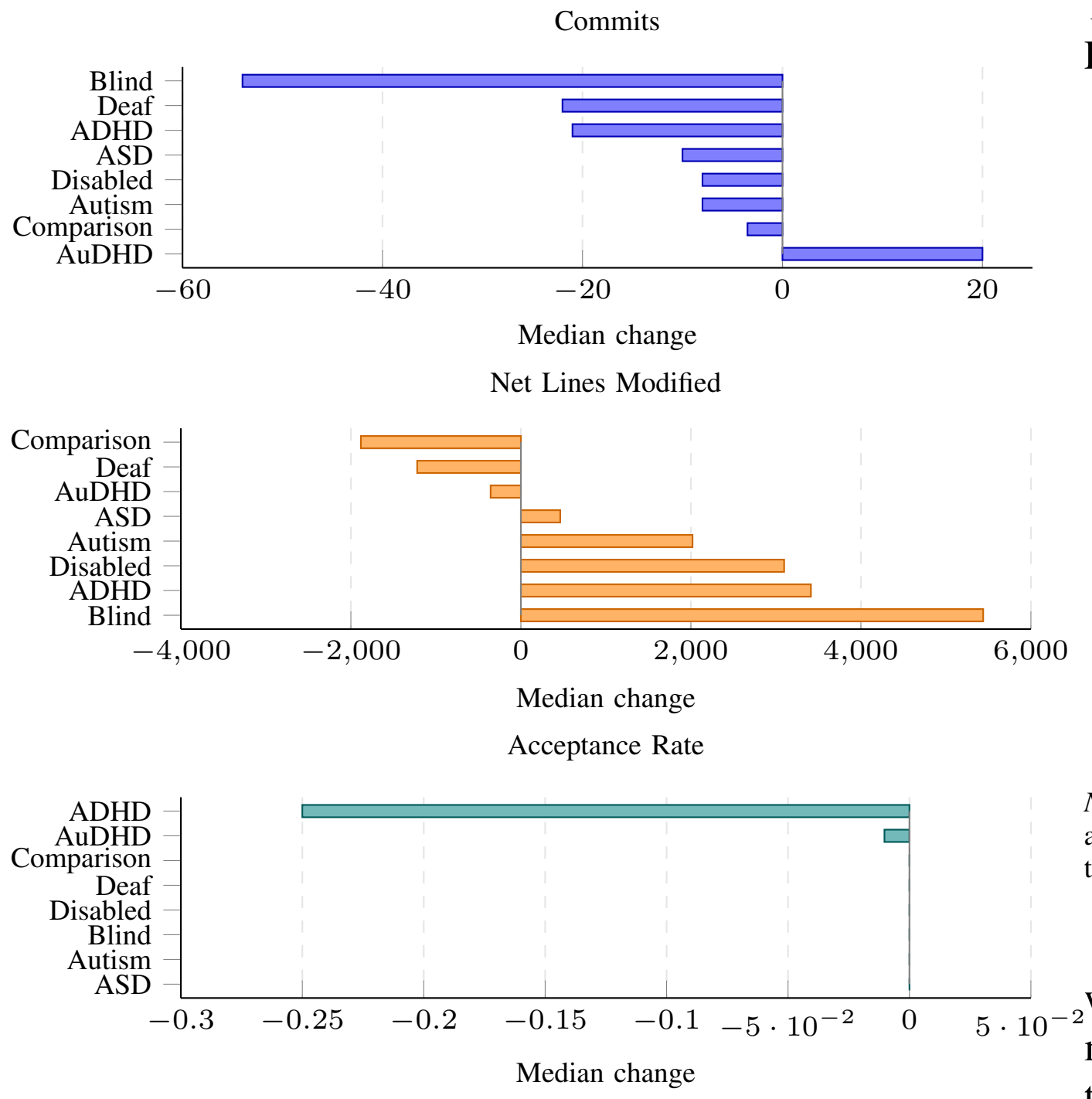
\begin{figure}[tb]
\centering
\footnotesize

\begin{tikzpicture}
\begin{axis}[
    width=\columnwidth,
    height=3.3cm,
    xbar,
    xmin=-60, xmax=25,
    bar width=3pt,
    enlarge y limits=0.08,
    xlabel={Median change},
    title={Commits},
    symbolic y coords={
        Blind,
        Deaf,
        ADHD,
        ASD,
        Disabled,
        Autism,
        Comparison,
        AuDHD
    },
    ytick=data,
    y dir=reverse,
    axis x line*=bottom,
    axis y line*=left,
    xmajorgrids=true,
    ymajorgrids=false,
    grid style={dashed,gray!20},
    tick label style={font=\scriptsize},
    label style={font=\scriptsize},
    title style={font=\scriptsize},
]
\addplot[
    fill=blue!50,
    draw=blue!70!black
] coordinates {
    (-54,Blind)
    (-22,Deaf)
    (-21,ADHD)
    (-10,ASD)
    (-8,Disabled)
    (-8,Autism)
    (-3.5,Comparison)
    (20,AuDHD)
};

\draw[black!50,thin]
    (axis cs:0,Blind) --
    (axis cs:0,AuDHD);

\end{axis}
\end{tikzpicture}

\vspace{0em}

\begin{tikzpicture}
\begin{axis}[
    width=\columnwidth,
    height=3.3cm,
    xbar,
    xmin=-4000, xmax=6000,
    bar width=3pt,
    enlarge y limits=0.08,
    xlabel={Median change},
    title={Net Lines Modified},
    symbolic y coords={
        Comparison,
        Deaf,
        AuDHD,
        ASD,
        Autism,
        Disabled,
        ADHD,
        Blind
    },
    ytick=data,
    y dir=reverse,
    axis x line*=bottom,
    axis y line*=left,
    xmajorgrids=true,
    ymajorgrids=false,
    grid style={dashed,gray!20},
    tick label style={font=\scriptsize},
    label style={font=\scriptsize},
    title style={font=\scriptsize},
]
\addplot[
    fill=orange!60,
    draw=orange!80!black
] coordinates {
    (-1882.5,Comparison)
    (-1220,Deaf)
    (-356,AuDHD)
    (463,ASD)
    (2019,Autism)
    (3097,Disabled)
    (3412,ADHD)
    (5438.5,Blind)
};

\draw[black!50,thin]
    (axis cs:0,Comparison) --
    (axis cs:0,Blind);

\end{axis}
\end{tikzpicture}

\vspace{0em}

\begin{tikzpicture}
\begin{axis}[
    width=\columnwidth,
    height=3.3cm,
    xbar,
    xmin=-0.3, xmax=0.05,
    bar width=3pt,
    enlarge y limits=0.08,
    xlabel={Median change},
    title={Acceptance Rate},
    symbolic y coords={
        ADHD,
        AuDHD,
        Comparison,
        Deaf,
        Disabled,
        Blind,
        Autism,
        ASD
    },
    ytick=data,
    y dir=reverse,
    axis x line*=bottom,
    axis y line*=left,
    xmajorgrids=true,
    ymajorgrids=false,
    grid style={dashed,gray!20},
    tick label style={font=\scriptsize},
    label style={font=\scriptsize},
    title style={font=\scriptsize},
]
\addplot[
    fill=teal!55,
    draw=teal!70!black
] coordinates {
    (-0.25,ADHD)
    (-0.0103,AuDHD)
    (0,Comparison)
    (0,Deaf)
    (0,Disabled)
    (0,Blind)
    (0,Autism)
    (0,ASD)
};

\draw[black!50,thin]
    (axis cs:0,ADHD) --
    (axis cs:0,ASD);

\end{axis}
\end{tikzpicture}

\caption{Median individual changes in productivity-related metrics
following the wider availability of GitHub Copilot. Groups are
ordered within each panel from the largest decrease to the largest
increase. Positive values indicate higher post-availability values,
whereas negative values indicate lower post-availability values.
Acceptance-rate changes are reported descriptively only.}
\label{fig:before_vs_after}
\end{figure}






\begin{table}[t]
\caption{Summary of Within-Group Activity Changes Following the Wider Availability of GitHub Copilot}
\label{tab:within_group_summary}
\centering
\footnotesize
\begin{tabular}{@{}llrrr@{}}
\toprule
\textbf{Group} & \textbf{Metric} & \textbf{$\Delta$} &
\textbf{$p_{\mathrm{FDR}}$} & \textbf{$|r|$} \\
\midrule
Deaf       & Commits   & -22     & 0.959 & 0.068 \\
           & Net Lines & -1220   & 1.000 & 0.029 \\
Disabled   & Commits   & -8      & 0.149 & 0.202 \\
           & Net Lines & +3097   & <0.001 & 0.367 \\
Blind      & Commits   & -54     & 0.714 & 0.310 \\
           & Net Lines & +5438.5 & 0.835 & 0.126 \\
ADHD       & Commits   & -21     & 0.219 & 0.251 \\
           & Net Lines & +3412   & 0.219 & 0.246 \\
Autism     & Commits   & -8      & 0.766 & 0.135 \\
           & Net Lines & +2019   & 0.766 & 0.163 \\
ASD        & Commits   & -10     & 0.321 & 0.413 \\
           & Net Lines & +463    & 1.000 & 0.004 \\
AuDHD      & Commits   & +20     & 0.835 & 0.117 \\
           & Net Lines & -356    & 0.766 & 0.220 \\
Comparison & Commits   & -3.5    & 0.766 & 0.110 \\
           & Net Lines & -1882.5 & 0.766 & 0.084 \\
\bottomrule
\end{tabular}

\vspace{2pt}
\parbox{\columnwidth}{\scriptsize
\textit{Note.} $\Delta$ represents the median individual change from
the pre-availability to the post-availability period.
$p_{\mathrm{FDR}}$ denotes the Benjamini--Hochberg adjusted
$p$-value, and $|r|$ is the absolute Wilcoxon effect size.
}
\end{table}

We next compared developers' productivity before and after the wider availability of GitHub Copilot. Figure~\ref{fig:before_vs_after} shows the median changes in commits, net lines modified, and pull request acceptance rates across disability groups\changed{, while Table~\ref{tab:within_group_summary} summarizes the statistical tests for commits and net lines modified. For pull request acceptance rate, as mentioned in Section~\ref{sec:min_group_size}, none of the disability groups met the minimum group size required for within-group statistical testing. We therefore report changes in pull request acceptance rate descriptively.}

\changed{After FDR correction (Table~\ref{tab:within_group_summary}), most changes did not remain statistically significant. The only statistically significant result after correction was an increase in net lines modified for developers who used the general ``Disabled'' label ($p_{\mathrm{FDR}} < .001$, $|r| = .367$). These results indicate that, following the wider availability of GitHub Copilot, developers in this group made fewer contributions while modifying more code during the post-availability period, even though their overall commit activity did not significantly increase.}

\changed{One possible explanation is that the wider availability of GitHub Copilot allowed developers with disabilities to complete larger amounts of code within a single contribution. Copilot may have reduced some of the effort involved in tasks such as generating or revising code, and enabled them to work on more extensive modifications without necessarily increasing how often they committed their work. Thus, the potential benefit may be reflected in supporting developers to accomplish more within individual contributions. It could also be that developers simply combined more work into fewer commits or worked on tasks that involved more code changes. \changedd{However, further investigation, particularly using self-reported or qualitative data, is needed to understand the factors behind this change. Previous studies suggest that Generative AI may influence how developers with disabilities work in ways that are not reflected in repository activity~\cite{cha2026gamechanger, Flores2025,Glazko2025}.}}


\changedd{The lack of statistical significance in some groups may reflect differences in sample size, as several disability groups contained relatively few developers. Although these changes were not statistically significant, their direction provides useful descriptive context as they may reveal patterns to understand how the different metrics changed relative to one another.} Descriptively, commit activity decreased in six of the seven disability groups (i.e., Deaf, Disabled, Blind, ADHD, Autism, ASD). However, five of these groups (i.e., Disabled, Blind, ADHD, Autism, ASD) showed an increase in net lines modified despite making fewer commits. In contrast, the comparison group showed decreases in both commits and net lines modified.

\changed{This broader pattern suggests that lower commit frequency did not consistently correspond to a lower amount of code modified among developers with disabilities. Although only the increase for the general ``Disabled'' group was statistically significant, the similar direction observed across several disability groups provides additional descriptive context for that result.}

\begin{center}
\fbox{
\parbox{0.96\columnwidth}{
\changed{\textbf{Finding 1.} Following the wider availability of GitHub Copilot, \textbf{most disability groups showed fewer commits but more code modified}.}}}
\end{center}

\subsection{Changes in Activity Levels}


The previous analysis examined changes in overall commit activity, but median changes may not capture how developers are distributed across different activity levels.
\changedd{The difference is important because Stray et al.~\cite{stray2026developerproductivitygithubcopilot} found that developers who used AI tools committed more and changed more code, but were already more active before the tools became available. We therefore examined how developers were distributed across different activity levels before and after the wider availability of GitHub Copilot.}
We grouped developers into three activity levels based on the number of commits in each period: fewer than 25 commits, 25-100 commits, and more than 100 commits. 

\changed{This exploratory analysis allows us to examine whether developers became more or less represented at different activity levels following the wider availability of GitHub Copilot. Because this analysis is descriptive, we focus on changes of at least 5 percentage points in the proportion of developers within an activity category. See Table~\ref{tab:activity_levels} for the distribution of developers across the three activity levels before and after the wider availability of GitHub Copilot.}







\begin{table}[t]
\centering
\caption{Distribution of developers across commit activity levels before and
after the availability of GitHub Copilot.}
\label{tab:activity_levels}
\resizebox{1\linewidth}{!}{%
\begin{tabular}{lrrrrrr}
\toprule
& \multicolumn{3}{c}{\textbf{Before}} & \multicolumn{3}{c}{\textbf{After}} \\
\cmidrule(lr){2-4} \cmidrule(lr){5-7}
\textbf{Disability group}
& $<25$
& $25$--$100$
& $>100$
& $<25$
& $25$--$100$
& $>100$ \\
\midrule

ADHD
& 10 & 26 & 25
& \cellcolor{green!15}21 $\uparrow$
& \cellcolor{red!15}15 $\downarrow$
& 25 \\

ASD
& 6 & 9 & 6
& \cellcolor{green!15}10 $\uparrow$
& \cellcolor{red!15}6 $\downarrow$
& 5 \\

AuDHD
& 4 & 5 & 6
& \cellcolor{red!15}3 $\downarrow$
& \cellcolor{red!15}4 $\downarrow$
& \cellcolor{green!15}8 $\uparrow$ \\

Autism
& 9 & 10 & 3
& 8 & 11 & 3 \\

Blind
& 2 & 4 & 8
& \cellcolor{green!15}3 $\uparrow$
& \cellcolor{red!15}3 $\downarrow$
& 8 \\

Deaf
& 3 & 8 & 2
& \cellcolor{green!15}4 $\uparrow$
& \cellcolor{red!15}4 $\downarrow$
& \cellcolor{green!15}5 $\uparrow$ \\

Disabled
& 7 & 12 & 26
& 7 & 14 & 24 \\

\bottomrule
\end{tabular}%
}

\vspace{2pt}
\parbox{\columnwidth}{\footnotesize
\textit{Note.} Highlighted cells indicate changes of at least 5 percentage points relative to the before period. Green indicates an increase and red indicates a decrease.
}
\end{table}

\changedd{Across the seven disability groups meeting our minimum sample size requirement of 10 developers,} five (i.e., ADHD, ASD, AuDHD, Blind, and Deaf) had fewer developers in the $25-100$ commits category after the wider availability of GitHub Copilot. For example, for ADHD, 26 developers were in the $25-100$ commits category before GitHub Copilot, but only 15 developers were in that category after. \changedd{This means that fewer developers in these five groups were represented in the moderate activity category after the wider availability of GitHub Copilot.} One possible explanation is that wider Copilot availability may have affected developers differently. Some may have become more active, while others may have remained less active, resulting in fewer developers at moderate activity levels.

\changed{However, where the activity increased instead differed across groups. For ADHD, ASD, and Blind, more developers were in the $<25$ commits category after wider Copilot availability. In contrast, for AuDHD, more developers were in the $>100$ commits category. Deaf developers showed a different pattern, with increases in both the $<25$ and $>100$ commits categories. Taken together, these patterns show that, after wider Copilot availability, some groups became more represented at lower activity levels, while others became more represented at higher activity levels or even at both ends of the activity range.}

\begin{center}
\fbox{
\parbox{0.96\columnwidth}{
\changed{\textbf{Finding 2.} Following the wider availability of GitHub Copilot, \textbf{activity became less centered around moderate levels} in most disability groups. \textbf{More developers moved to lower activity levels and others to higher levels}, leaving fewer developers in the middle.
}}}
\end{center}




\subsection{Differences Across Disability Groups}

We next examined whether changes in productivity differed across disability groups. \changed{For each developer, we calculated a change score for each metric by subtracting the pre-availability value from the post-availability value, and then compared the distribution of these change scores across disability groups.} This comparison is important because different disabilities may involve different accessibility needs and may therefore interact differently with AI tools. \changed{We first compared the magnitude of change across the broader Physical, Non-Physical, Visual, Disabled, and comparison groups (i.e., developers with no identified disability disclosure). We then compared the magnitude of change between disability subtypes within the Non-Physical category (i.e., ADHD, Autism, ASD, and AuDHD). This was the only broader category containing multiple disability subtypes that met our minimum sample-size requirement of 10 users.}

\begin{table}[t]
\caption{\changed{Comparison of Changes Across Broader Groups and Non-Physical Subtypes}}
\label{tab:group_comparisons}
\centering
\footnotesize

\begin{tabular}{@{}llccc@{}}
\toprule
\textbf{Analysis} & \textbf{Metric} & \textbf{$H$} & \textbf{$p$} & \textbf{$p_{\mathrm{FDR}}$} \\
\midrule

\multirow{2}{*}{Broader groups}
& $\Delta$ Commits   & 5.495 & 0.240 & - \\
& $\Delta$ Net Lines & 6.450 & 0.168 & - \\
\midrule

\multirow{2}{*}{Non-Physical subtypes}
& $\Delta$ Commits   & 2.932 & 0.402 & 0.728 \\
& $\Delta$ Net Lines & 1.304 & 0.728 & 0.728 \\

\bottomrule
\end{tabular}

\vspace{2pt}
\parbox{\columnwidth}{\scriptsize
\textit{Note.} Broader-group comparisons include the Physical,
Non-Physical, Visual, Disabled, and comparison groups.
Non-Physical subtype comparisons include ADHD, Autism, ASD,
and AuDHD. Pull request acceptance rate was not statistically
tested because no group met the minimum eligible sample size.
}
\end{table}

\changed{As shown in Table~\ref{tab:group_comparisons}, the $H$ statistic reflects how much the change scores differed across the broader groups, while the $p$-value indicates whether these differences were statistically significant. We found no significant differences across the groups for either commits or net lines modified. Overall, there was no statistically supported difference in how the Physical, Non-Physical, Visual, Disabled, and comparison groups changed from before to after the wider availability of GitHub Copilot.}

\changed{As for comparison between disability subtypes within the Non-Physical (see Table~\ref{tab:group_comparisons}), we found no statistically significant differences in the magnitude of change. This was consistent for both commits and net lines modified, with all corrected $p$-values remaining non-significant. This means ADHD, Autism, ASD, and AuDHD did not differ significantly in how their commits or net lines modified changed from before to after wider Copilot availability.}



\begin{center}
\fbox{
\parbox{0.96\columnwidth}{
\textbf{Finding 3.} \textbf{No statistically supported differences were found in how commits or net lines modified changed across disability groups} following the wider availability of GitHub Copilot.}}
\end{center}

\section{Discussion} \label{sec:Discussion}

\subsection{Limited Productivity Changes}

Overall, the findings do not show a clear increase in observable productivity among self-identified disabled developers following the wider availability of GitHub Copilot. Changes varied across productivity metrics and disability groups, but most were not statistically significant after FDR correction. In addition, the magnitude of these changes did not differ significantly across disability groups. \changed{We argue that the limited changes observed in this study do not reflect a complete absence of AI use, as explicit AI-related mentions were observed at broadly similar proportions in the disability and comparison samples, as explained in \ref{visibility}.}

The only statistically significant change after FDR
correction was an increase in net lines modified for developers
who used the general ``Disabled'' label. This increase was not
accompanied by a statistically significant increase in commits. More broadly, the descriptive results showed that five of the
seven disability groups had fewer commits but more net lines
modified. This suggests that contribution frequency and the
amount of code modified did not necessarily change in the same
direction following the wider availability of GitHub Copilot.

These findings add nuance to previous research on
Generative AI and developer productivity. Stray et al.~\cite{stray2026developerproductivitygithubcopilot} found that developers reported feeling more productive when using Copilot, while their repository activity did not increase significantly after adoption. Other studies have reported benefits such as faster task completion and improved perceived productivity~\cite{peng2023the,10.1145/3520312.3534864}. \changed{Our results suggest that
the absence of an increase in repository activity does not
necessarily mean that developers' work remained unchanged. }

\changed{Together, these findings suggest that improvements associated with AI may not necessarily result in an overall increase in activity on GitHub. Our analysis adds further nuance to this result. Although overall changes were limited, developers were distributed differently across activity
levels after the wider availability of GitHub Copilot. In most disability groups, fewer developers were represented at moderate
activity levels, while some groups became more represented at lower activity levels and others at higher activity levels.}



\subsection{What GitHub Metrics May Miss for Developers with Disabilities}
For developers with disabilities, the benefits of AI may not necessarily appear as more commits, more lines modified, or higher pull request acceptance rates. Instead, AI may influence how development work is carried out, for example by reducing the effort needed to complete tasks, making information easier to access, or helping developers manage barriers that arise during software development.

Previous studies on visually impaired developers illustrate this distinction. Flores-Saviaga et al.~\cite{Flores2025} found that Generative AI coding assistants could increase developers' sense of control and support higher-level reasoning, while Cha et al.~\cite{cha2026gamechanger} reported reduced access labor and less reliance on sighted colleagues. These benefits relate more to how developers carry out their work than to how much activity they produce.

A similar distinction can be seen for neurodivergent developers. Gama et al.~\cite{gama2025sociotechnical} found that developers with ADHD and ASD can experience challenges related to task initiation, organization, focus, and context switching. Glazko et al.~\cite{Glazko2025} further showed that Generative AI can support neurodivergent users with tasks such as accessing information, organizing thoughts, and managing executive dysfunction. These kinds of changes may make development work easier without necessarily resulting in more commits or lines modified.

The same may also apply to Deaf and Hard-of-Hearing users. Huffman et al.~\cite{huffman2025weuseithearing} found that Generative AI could support aspects of communication and learning, although experiences differed depending on users' language preferences and needs. These benefits are more closely related to accessibility and ease of interaction.

\changed{Taken together, our results do not show a broad increase
or decrease in activity following the wider
availability of GitHub Copilot. At the same time, the descriptive
activity-level analysis shows that this overall stability does not
mean that all developers followed the same pattern. It may be that some disability
groups became more represented at lower activity levels, while
others became more represented at higher levels. Our repository
data cannot determine whether these changes reflect benefits from
Generative AI, such as reduced effort or improved accessibility.
Such effects may be more subtle and may not translate directly into
more repository activity.}

\subsection{Implications}
Our findings suggest that studying the effects of Generative AI on disabled developer populations requires more than productivity metrics alone. Changes in how work is performed may be just as important as changes in visible output. Research on the influence of AI on disabled developers should therefore consider both observable activity and developers' experiences of effort, accessibility, independence, and ease of completing development tasks.

Future studies could combine repository mining with surveys and/or interviews to capture these different dimensions. Repository data can show whether activity changes over time, while self-reported or qualitative data can help explain whether Generative AI changes how developers approach tasks, manage accessibility barriers, or experience their work. This would provide a more complete picture of how AI tools affect developers with disabilities and other underrepresented groups.

\section{Threats to Validity}

A first threat concerns the identification of developers with disabilities. Our keyword-based approach does not cover the full range of disabilities and is limited to English terminology. Developers who used
different terminology may not have been identified. \changed{For example,
ADHD is commonly abbreviated as ``ADHS'' in German or ``TDAH'' in Spanish, whereas our search relied on the English
term ``ADHD.''} To mitigate this threat, we curated the keyword
list based on existing definitions, manually reviewed developers'
profile descriptions to confirm that disability-related terms
referred to self-identification, and removed ambiguous keywords
when validation showed that they produced unreliable results.



Our reliance on publicly self-disclosed disability information also introduces selection bias. We only include developers who chose to disclose a disability in their GitHub profile, and therefore our sample should not be considered representative of all developers with disabilities. 
We therefore limit our claims to developers who publicly self-identify as having a disability. 

A further threat concerns identifying Generative AI use. Developers are not required to report AI assistance in commit messages, meaning that explicit mentions are likely to capture only a subset of actual use. \changed{To reduce under-detection, we searched for
multiple Generative AI tools as well as general AI-related terms
rather than GitHub Copilot alone. However, we do not interpret
the absence of such mentions as evidence that a developer did not
use Generative AI. We therefore use these mentions only to assess
whether Generative AI use is explicitly visible in repository
activity.}

The before-and-after design also does not establish that individual developers adopted GitHub Copilot when it became freely available. To address this, we use 18 December 2024 only as a point of wider availability and avoid interpreting the observed changes as causal effects of Copilot adoption. 

Another threat concerns how productivity is measured. Repository metrics such as commits, net lines modified, and pull request acceptance capture only observable aspects of development activity. To reduce reliance on a single measure, we considered multiple productivity metrics and additionally examined changes across low, medium, and high commit activity levels. Nevertheless, these measures cannot capture aspects such as cognitive effort, task complexity, perceived productivity, or accessibility-related experiences.


\section{Conclusions and Future Work} \label{sec:Conclusions}


This study investigated whether the wider availability of GitHub Copilot was associated with changes in observable productivity-related activity among GitHub users who publicly self-identify as having a disability. Using publicly available profile information, we identified developers across multiple disability groups and compared their repository activity before and after GitHub Copilot became freely available to all GitHub users.
Overall, we observed several descriptive changes in development activity, but did not detect a broad statistically significant increase or decrease across disability groups. After FDR correction, the only statistically
significant within-group change was an increase in net lines modified for developers using the general ``Disabled'' label. However, several disability groups showed fewer commits alongside
more code modified, while the activity-level analysis showed that
changes were not distributed uniformly across low, moderate, and
high activity levels. We also found no statistically significant
differences in the magnitude of change across disability groups.
These findings suggest that changes, if any, following the wider availability of GitHub Copilot may not be captured by increases in repository activity alone. Repository metrics provide one view of developer activity, but they cannot show whether Generative AI reduces effort, improves accessibility, or otherwise changes how development work is carried out. 

Future studies should therefore combine repository measures with developers' self-reported or qualitative experiences. 
Because changes were not distributed uniformly across low, moderate, and high activity levels, future research should examine how baseline engagement moderates AI utility. For example, occasional contributors may use AI to overcome initial onboarding barriers in unfamiliar codebases, whereas high-activity maintainers might rely on it for routine maintenance or documentation. Stratified longitudinal modeling could clarify who benefits most and under what task conditions.
Moreover, it would be interesting for companies to replicate our study in their industrial closed-source repositories. Organizational case studies could also enable the use opt-in IDE telemetry or to compare verified AI users against non-users and measure how specific features (e.g., inline completions vs. conversational chat) are utilized by different groups.

\section{Data Availability} The data is available \href{https://figshare.com/s/a05fd8d8e601d7cb6dcb}{in our supplementary material}.


\bibliographystyle{IEEEtran}
\bibliography{IEEEabrv,references}
\end{document}